\documentclass[5p,times,twocolumn]{elsarticle}

\usepackage{amsmath}
\usepackage{amssymb}
\biboptions{comma,sort&compress}
\usepackage[utf8]{inputenc}
\usepackage{amsmath}
\usepackage{amssymb}
\usepackage{graphicx}
\usepackage{dcolumn}
\usepackage{bm}
\usepackage{hyperref} 
\hypersetup{
colorlinks = true,
linkcolor = blue,
anchorcolor = blue,
citecolor = blue,
filecolor = blue,
urlcolor = blue
}
\usepackage{float}
\usepackage{color}
\usepackage{ulem}

\journal{Physics Letters B}

\begin{document}

\begin{frontmatter}

\title{Hydrodynamics from free-streaming to thermalization and back again}

\author{Chandrodoy Chattopadhyay\corref{cor1}}
\ead{chattopadhyay.31@osu.edu}
\cortext[cor1]{Corresponding author}
\author{Ulrich Heinz}
\ead{heinz.9@osu.edu}
\address{Department of Physics, The Ohio State University, Columbus, Ohio 43210-1117, USA}


\begin{abstract} 

We study the evolution of the Knudsen and Reynolds numbers in (0+1)-dimensionally expanding fluids with Bjorken symmetry for systems whose microscopic mean free path rises more quickly with time than usually assumed. This allows us to explore within a simple 1-dimensional model the transition from initially thermalizing to ultimately decoupling dynamics. In all cases studied the dynamics is found to be controlled by a hydrodynamic attractor for the Reynolds number whose trajectory as a function of Knudsen number makes a characteristic turn as the dynamics changes from thermalizing to decoupling. We argue that this feature is robust and should also manifest itself in realistic 3-dimensional simulations of  expanding heavy-ion collision fireballs.   

\end{abstract}

\begin{keyword}
relativistic hydrodynamics \sep heavy-ion collisions \sep quark-gluon plasma
\end{keyword}

\end{frontmatter}



\textit{\textbf{1. Introduction.}} 
%
The impressive success of modern formulations of relativistic dissipative hydrodynamics in explaining macroscopic properties of matter formed in high-energy collisions ranging from proton-on-proton to lead-on-lead \cite{Romatschke:2007mq, Song:2007ux, Song:2008si, Schenke:2010nt, Heinz:2013th} has raised interest in the apparently simple question: When does a macroscopic system start to exhibit hydrodynamic behavior? Not only has the answer to this question recently witnessed a paradigm shift \cite{Heller:2011ju, Heller:2013fn,  Heller:2015dha, Kurkela:2015qoa, Blaizot:2017lht, Romatschke:2017vte, Spalinski:2017mel,  Strickland:2017kux, Romatschke:2017acs, Behtash:2017wqg, Blaizot:2017ucy, Kurkela:2018wud, Mazeliauskas:2018yef,  Behtash:2019txb, Heinz:2019dbd, Blaizot:2019scw} but so has the notion of what constitutes a `hydrodynamic' theory. Traditionally, relativistic dissipative hydrodynamics has been regarded as an effective theory formulated as an expansion in gradients of macroscopic degrees of freedom (densities and flow velocities). In this approach applicability of fluid dynamics implies proximity to local thermodynamic equilibrium: large gradients generate large negative contributions to the local entropy density \cite{Loganayagam:2008is, Chattopadhyay:2014lya}, thus invalidating both convergence of the gradient expansion and proximity to local equilibrium. 

This traditional approach has multiple problems. At first order in gradients (relativistic Navier-Stokes theory) the theory is acausal and unstable \cite{Hiscock:1983zz, Hiscock:1985zz}. Adding higher-order gradient corrections provides no systematic improvement since it was recently discovered that the gradient expansion is plagued by factorially growing coefficients and generically diverges \cite{Heller:2013fn}. The acausality problem can be circumvented by promoting the dissipative currents to independent dynamical degrees of freedom. These obey relaxation-type equations involving higher-order gradient terms multiplied by transport coefficients \cite{Israel:1976tn, Israel:1979wp} that reflect the interplay between microscopic interactions facilitating thermalization and gradient-driven macroscopic expansion which drives the system away from thermal equilibrium \cite{Kurkela:2011ub}.  The structure of these relaxation-type equations is universal and restricted only by macroscopic symmetries while the transport coefficients (which control the relative importance of different gradient terms) are medium properties reflecting microscopic interaction dynamics \cite{Baier:2007ix, Denicol:2012cn, Heinz:2015gka, McNelis:2018jho}.

Pursuing this program has led to the development of a variety of so-called second- and third-order causal dissipative hydrodynamic theories \cite{Muronga:2001zk, Baier:2007ix, Denicol:2012cn, Jaiswal:2013npa, Jaiswal:2013vta, Grozdanov:2015kqa}, including an anisotropic hydrodynamic framework that addresses the particularly large discrepancy between longitudinal and transverse flow gradients during the early stages of relativistic heavy-ion collisions \cite{Martinez:2010sc, Florkowski:2010cf, Bazow:2013ifa, Molnar:2016vvu, McNelis:2018jho} and effectively resums certain classes of gradients to infinite order \cite{Strickland:2014pga, Alqahtani:2017mhy}. In a highly symmetric limit, Bjorken flow \cite{Bjorken:1982qr}, which applies to the earliest expansion stage of the medium created in such collisions, it was found that the diverging gradient series can be resummed in its entirety using Borel resummation \cite{Heller:2015dha}, resulting in a time evolution that agrees with the above-mentioned low-order causal hydrodynamic approaches at late times, when the system approaches local thermal equilibrium, but remains well-behaved even at very early times when the system is very far away from local thermal equilibrium \cite{Romatschke:2017vte,  Florkowski:2017olj}. Both the Borel-resummed full gradient expansion and the low-order causal hydrodynamic theories exhibit attracting behavior, in the sense that for arbitrary off-equilibrium initial conditions the dynamics  quickly settles on the hydrodynamical attractor, on a time scale given by the microscopic relaxation time or shorter \cite{Romatschke:2017vte, Jaiswal:2019cju, Kurkela:2019set}, and then follows it over macroscopic time scales until deviations from local equilibrium eventually vanish and the system becomes an ideal fluid.

Qualitatively, the hydrodynamic attractors associated with these partially or fully resummed hydrodynamic theories are all very similar \cite{Romatschke:2017vte, Jaiswal:2019cju} but differ in detail depending on the underlying microscopic dynamics and the approximations made when coarse-graining it to obtain a macroscopic hydrodynamic description. For systems whose microscopic dynamics can be described by classical kinetic theory using the relativistic Boltzmann equation it was found that the evolution of non-hydrodynamic moments of the distribution function, and of that function itself, is also controlled by attractors \cite{Strickland:2018ayk}, and that both anisotropic \cite{Behtash:2017wqg, Martinez:2017ibh, Strickland:2017kux} and third-order Chapman-Enskog hydrodynamics \cite{Jaiswal:2013vta} describe this kinetic attractor with precision for both Bjorken \cite{Bjorken:1982qr} and Gubser \cite{Gubser:2010ze} flows. The latter observation is of particular importance since, in contrast to the purely 1-dimensional longitudinal expansion of Bjorken flow, Gubser flow is intrinsically 3-dimensional, with such rapid transverse expansion that the system actually moves away from local equilibrium and becomes asymptotically free-streaming \cite{Denicol:2014xca, Denicol:2014tha}. Still, anisotropic and 3$^\mathrm{rd}$-order hydrodynamics describe this evolution accurately even in the late-time free-streaming limit \cite{Chattopadhyay:2018apf}. Existence of the attractor renders the asymptotic dynamical evolution insensitive to initial conditions -- memory of the initial state is lost on the microscopic scattering time scale or even earlier \cite{Heller:2016rtz, Heller:2018qvh, Jaiswal:2019cju, Kurkela:2019set}.

These prior observations have established that, contrary to previous belief, dissipative relativistic fluid dynamics can provide a quantitatively precise description of the far-off-equilibrium evolution of very rapidly and anisotropically expanding systems controlled by large space-time gradients, owing to the existence of off-equilibrium hydrodynamic attractors. However, most studies so far invoked simple flows with a high degree of symmetry, begging the question whether the insights gained are more universally applicable and can perhaps explain the phenomenological success of (3+1)-d numerical fluid dynamics in describing heavy-ion collisions. First numerical studies of attractor behavior for (3+1)-d systems with longitudinal boost invariance \cite{Romatschke:2017acs, Kurkela:2018qeb, Kurkela:2019kip, Kurkela:2019set} established the persistence of attractive behavior also in these more general systems where the expansion evolves from being primarily 1-dimensional at early times to full-fledged 3-dimensional expansion where the Knudsen number (i.e. the ratio between microscopic and macroscopic length scales) grows again at late times. However, the nature of these late-time attractors in 3+1 dimensions and their dynamical emergence from the early-time Bjorken attractor remain incompletely understood.        

\textit{\textbf{2. Modeling the transition from thermalizing to decoupling dynamics in Bjorken flow.}}
%
The key feature distinguishing realistic (3+1)-d expansion from (0+1)-d Bjorken expansion is a rapidly increasing expansion rate for the former as the effective dimensionality of the expansion increases with time \cite{Kolb:2003dz, Song:2007ux}. The physics controlling the microscopic scattering, on the other hand, remains unchanged, resulting for approximately massless constituents in a temperature dependence of the relaxation time $\tau_\pi(T)$ given by $T(\tau)\tau_\pi(\tau)=5\bar\eta \approx \text{const.}$ ($\bar\eta =\eta/s$ is the shear viscosity to entropy density ratio). In Bjorken flow we cannot change the expansion rate $\theta \equiv \partial{\cdot}u=1/\tau$ (where $u^\mu$ is the flow velocity field and $\tau$ is the longitudinal proper time \cite{Bjorken:1982qr}). However, we can change the time dependence of the Knudsen number Kn${\,=\,}\theta\,\tau_\pi{\,=\,}\tau_\pi/\tau$, making it grow at late times just as in the realistic (3+1)-d simulations: We simply change the temperature dependence of $\tau_\pi$ to $\tau_\pi = (5\bar\eta/T_0) (T_0/T)^{\Delta(\tau)}$, with the power $\Delta(\tau)$ increasing from $\Delta=1$ at early times to $\Delta > 3$ at late times.\footnote{%
 	Alternatively, one could change the structure of the collision term, see \cite{Denicol:2019lio}.}
This allows us to study the evolution of the system and the hydrodynamic attractor as its dynamics changes character, from moving towards local equilibrium at early times to running away from it at late times when it approaches free-streaming. To set the stage, we first consider hydrodynamic evolution for several gradually increasing constant values of $\Delta$ before moving to the interesting time-dependent case $\Delta(\tau)$. 

Earlier work \cite{Martinez:2017ibh, Behtash:2017wqg, Chattopadhyay:2018apf} has established anisotropic (aHydro) and 3$^\mathrm{rd}$-order Chapman-Enskog (CE) hydrodynamics as particularly accurate approximations to an underlying kinetic theory of massless degrees of freedom,\footnote{%
	Although the precision of these hydrodynamic approaches for Bjorken flow has been shown
	only for conformal theories where $T\tau_\pi{\,=\,}$const. we have no reason to doubt that 
	it persists for more general temperature dependences of $\tau_\pi$. 
	\label{fn1}}
and we will hence focus on these two specific frameworks. Even though a nonlinear temperature dependence of $1/\tau_\pi$ breaks conformal symmetry we continue to ignore bulk viscosity and set the bulk viscous pressure to $\Pi{\,=\,}0$. Symmetries of Bjorken flow limit the shear stress $\pi^{\mu\nu}$ to a single independent component for which we take the unitless ratio $ \bar\pi \equiv -\tau^2\pi^{\eta_s\eta_s}/(\epsilon{+}P)$; here $\eta_s$ denotes space-time rapidity along the beam direction and $\epsilon$ and $P(\epsilon)$ are the fluid's energy density and equilibrium pressure. We use a conformal equation of state $P{\,=\,}\epsilon/3$ such that $\epsilon{+}P{\,=\,}\frac{4}{3}\epsilon$, $s{\,\propto\,}\epsilon^{3/4}$, and $T{\,\propto\,}\epsilon^{1/4}$. We also introduce the shorthand $\bar{\tau}{\,=\,}\tau/\tau_\pi$ for the longitudinal proper time in units of the microscopic relaxation time $\tau_\pi$. Subscripts 0 denote initial conditions.

The evolution follows the energy conservation law \cite{Molnar:2016gwq}
\begin{equation}
\label{econs}
   \tau\frac{d\ln[(\epsilon/\epsilon_0)^{3/4}]}{d\tau} = -(1- \bar{\pi})
\end{equation}
together with the shear stress relaxation equation
\begin{equation}
\label{ahydro}   
   \tau \frac{d\bar{\pi}}{d\tau} + \bar\pi \bar\tau = 
   \frac{5}{12} -  \frac{4}{3} \bar\pi - \frac{4}{3}\bar\pi^2 - \frac{3}{4} {\cal F}(\bar\pi)
\end{equation}
for anisotropic hydrodynamics \cite{Molnar:2016gwq, Martinez:2017ibh} or instead
\begin{equation}
\label{CE}
   \tau \frac{d\bar{\pi}}{d\tau} + \bar{\pi}\bar{\tau}  =  
   \frac{4}{15} - \frac{10}{21} \bar{\pi} - \frac{412}{147} \bar{\pi}^2 
\end{equation} 
for 3$^\mathrm{rd}$-order Chapman-Enskog hydrodynamics \cite{Jaiswal:2013vta}. The function ${\cal F}(\bar\pi)$ in Eq.\,(\ref{ahydro}) was originally defined in Eq.\,(36) of Ref.\,\cite{Martinez:2017ibh} in terms of the momentum deformation parameter of an underlying anisotropic distribution function (see also \cite{Strickland:2017kux}). This definition is restricted to the interval $-\frac{1}{2}{\,\leq\,}\bar\pi{\,\leq\,}\frac{1}{4}$ (i.e. to positive longitudinal and transverse effective pressures) in which ${\cal F}(\bar\pi)$ smoothly interpolates between ${\cal F}(-\frac{1}{2}){\,=\,}1$ and ${\cal F}(\frac{1}{4}){\,=\,}0$. Here we use a polynomial fit\footnote{%
	For $\bar\pi{\,\leq\,}1/3$ we use ${\cal F}(\bar\pi)=0.2 - 1.18526 \bar\pi +1.30385 \bar{\pi}^2
	+0.948719 \bar{\pi}^3$; for $\bar\pi{\,>\,}1/3$ we set ${\cal F}(\bar\pi)=-0.015076$.}
to extrapolate ${\cal F}(\bar\pi)$ beyond this range and thereby Eq.~(\ref{ahydro}) into transient regions of negative transverse or longitudinal pressures.

%
\begin{figure*}[htb!]
\centering
  \includegraphics[width=0.8\textwidth]{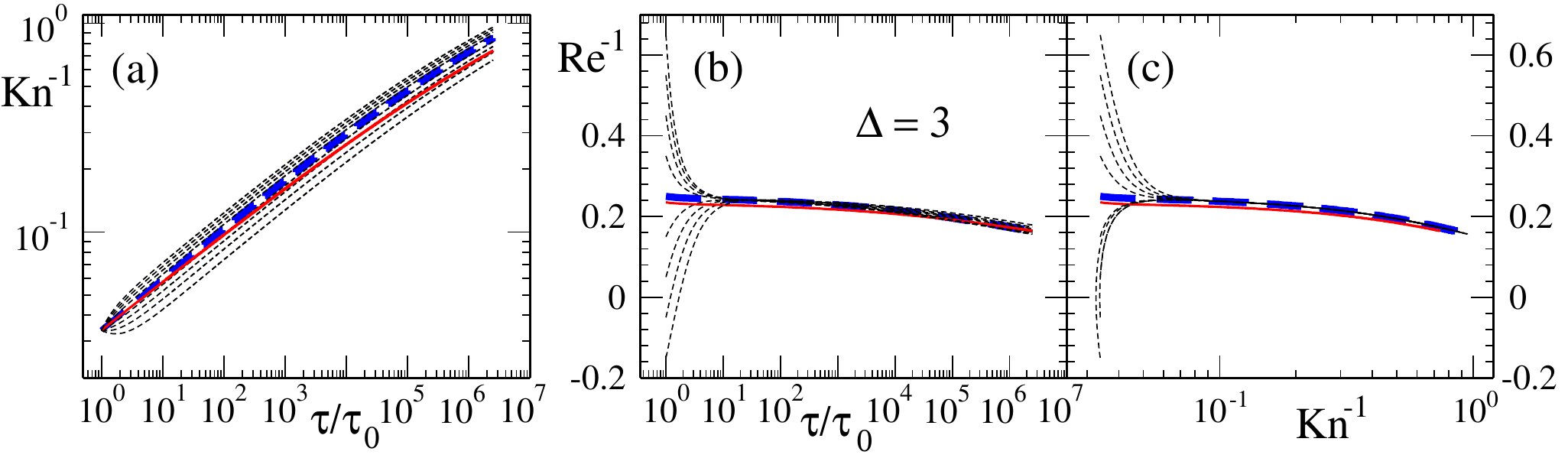}
 \vspace{-3mm}
 \caption{(Color online) Evolution for $\Delta=3$ of the inverse Knudsen and Reynolds 
 	numbers as functions of $\tau/\tau_0$ (panels a and b) and relative to each other (panel c).
	The thick blue and thin black dashed lines are obtained using aHydro whereas the red 
	solid line is the attractor for the third-order Chapman-Enskog theory. In panel c 
	the arrow of time points from left to right.
	\label{F1}
	\vspace*{-4mm}}
\end{figure*} 
%

The two key parameters describing the off-equilibrium evolution (\ref{econs}-\ref{CE}) are the inverse Knudsen number  
\begin{equation}
\label{Kn}
  \mathrm{Kn}^{-1} = \bar\tau 
  = \bar{\tau}_0 \left( \frac{\tau}{\tau_0} \right) \left( \frac{T}{T_0} \right)^{\Delta} 
  \qquad \left(\text{with}\ \bar\tau_0{\,=\,}\frac{\tau_0 T_0}{5 \bar{\eta}}\right)
\nonumber
\end{equation}
and the inverse Reynolds number Re$^{-1}=\bar\pi$. Kn drives deviations from local equilibrium while Re$^{-1}$ quantifies the fluid's dissipative response in terms of its actual deviation from local equilibrium. For Bjorken flow the evolution of the inverse Knudsen number is given by 
\begin{equation}
\label{Knudsen}
   \frac{\tau}{\bar\tau}\frac{d\bar{\tau}}{d\tau} = 1 - \frac{\Delta}{3}(1{-}\bar{\pi})
   + \tau \frac{d \Delta}{d\tau} \, \ln(T/T_0).
\end{equation}
For $\Delta{\,=\,}$const., Eqs.\,(\ref{ahydro})-(\ref{Knudsen}) show that, for identical initializations $(\bar\tau_0, \bar\pi_0)$ at different starting times $\tau_0$, the evolutions of both the inverse Knudsen and inverse Reynolds numbers are functions only of the scaled time $\ln(\tau/\tau_0)$. The attractor for $\bar\pi$ is obtained by evolving (\ref{ahydro},\ref{CE}) with initial conditions at $\tau_0{\,=\,}0$ corresponding to zero right hand sides, i.e.\ with $\bar\pi^\mathrm{attr}_{0,\mathrm{aHydro}}{\,=\,}\frac{1}{4}$ and $\bar\pi^\mathrm{attr}_{0,\mathrm{CE}} \approx 0.235$ \cite{Jaiswal:2019cju}, respectively. The first value agrees with the free-streaming limit; the second is ${\sim\,}1\%$ off. We set $T_0{\,=\,}2$\,GeV at $\tau_0{\,=\,}0.004$\,fm/$c$ and $\bar\eta = 0.24 \approx 3/(4\pi)$. 

\begin{figure}[htbp!]
\begin{center}
 \includegraphics[width=\linewidth]{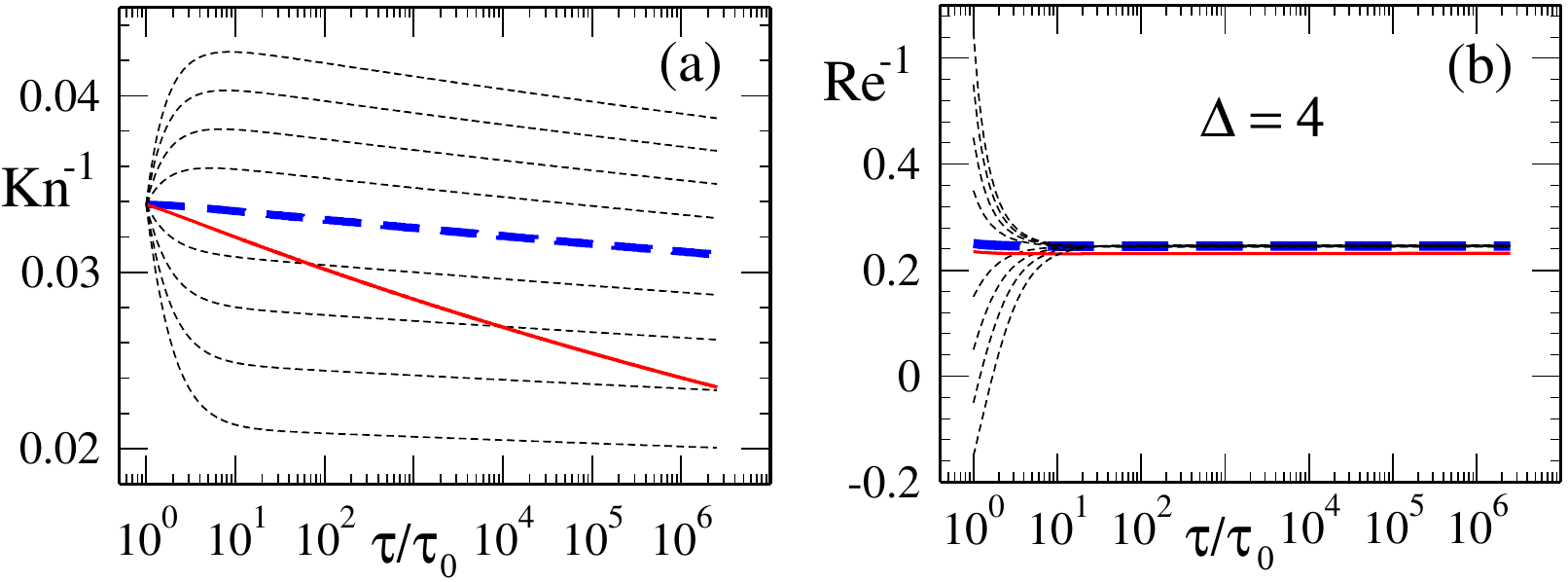}
\end{center}
 \vspace{-.5cm}
  \caption{
	Similar to Fig.~\ref{F1} (without panel c) but for $\Delta=4$.
	\label{F2}
	\vspace*{-2mm}}
\end{figure}

\begin{figure*}[htb!]
\centering
\vspace*{-3mm}
 \includegraphics[width=0.8\textwidth]{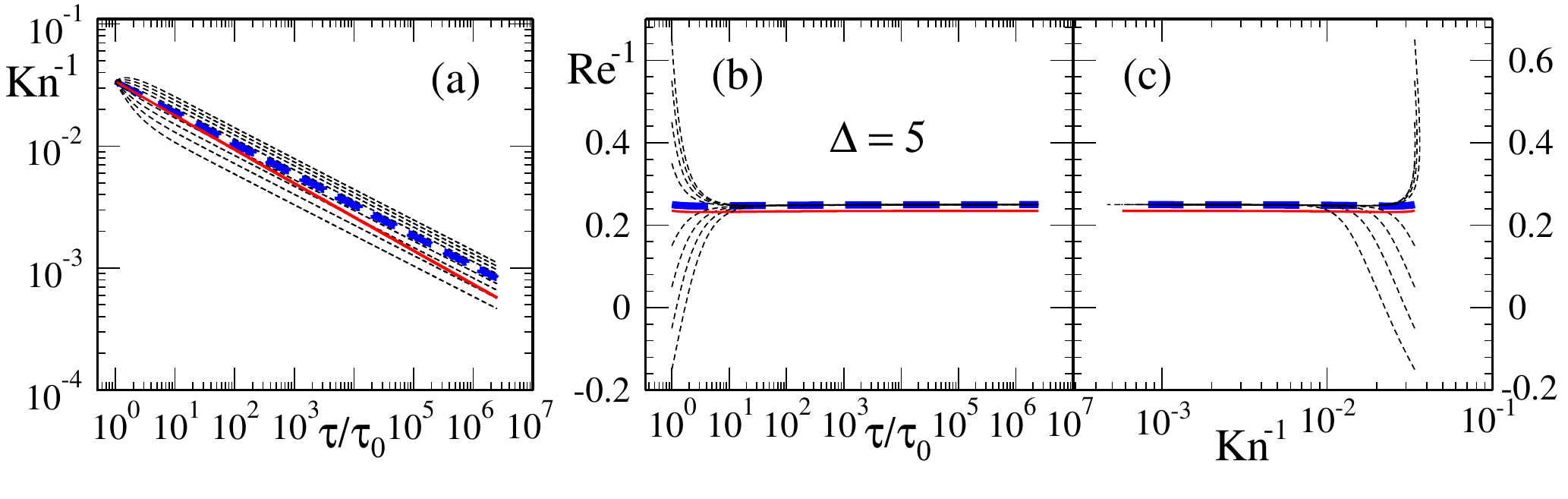}
 \vspace{-3mm}
  \caption{(Color online) 
  	Same as Fig.~\ref{F1} but for $\Delta = 5$. In panel c the arrow of time now points 
	from right to left.
	\label{F3}
	\vspace*{-3mm}}
\end{figure*}

\textit{\textbf{3.\,Results.}}{\it\ (a)\,Constant $\Delta$.}
%
We solve Eqs.~(\ref{ahydro})-(\ref{Knudsen}) numerically for different choices of $\Delta$, covering a wide range of far-off-equilibrium initial conditions with large fixed initial Knudsen number. Previous studies \cite{Heller:2016rtz} have shown (see also Fig.~\ref{F4} below) that for $\Delta{\,<\,}3$ the inverse Knudsen number grows with time (i.e. the system approaches local equilibrium) as $\bar\tau{\,\sim\,}\tau T_\mathrm{id}^\Delta{\,\sim\,}\tau^{1{-}\Delta/3}$ where $T_\mathrm{id}(\tau)$ is the ideal Bjorken cooling law. Hence we start our presentation of results with the more interesting choices $\Delta{\,=\,}3$ (Fig.~\ref{F1}), $\Delta{\,=\,}4$ (Fig.~\ref{F2}), and $\Delta{\,=\,}5$ (Fig.~\ref{F3}) for which we do not expect the system to thermalize.

Figs.~\ref{F1}a,b show the time evolution with aHydro of the inverse Knudsen and Reynolds numbers  for a variety of initial values $\bar\pi_0$. Fig.~\ref{F1}c shows the evolution of the inverse Reynolds number as a function of inverse Knudsen number. In all three panels the thick blue dashed line is the aHydro attractor, obtained by solving the equations with the above-mentioned attractor initial conditions implemented at $\tau_0{\,\approx\,}0$. The thin solid red line shows the corresponding CE attractor for comparison; it lies slightly below the aHydro attractor, indicating a somewhat smaller shear stress and correspondingly reduced viscous heating throughout the evolution. 

The black dashed lines in Fig.~\ref{F1}a correspond to different initial conditions $\bar\pi_0$. The splitting between them is caused by variations in viscous heating at early times before the differing initial deviations from the attractor have decayed. After their decay, all dashed black lines are parallel, i.e. they all follow the same power law.

According to Figs.~\ref{F1}b,c, for $\Delta{\,=\,}3$ the inverse Reynolds number remains approximately constant for $\tau{\,\lesssim\,}10^3\,\tau_0$, decaying to zero long after the interval shown in the plot.\footnote{%
	Contrary to naive first expectation, the system does eventually thermalize into local equilibrium,
	due to shear-stress-induced viscous heating as reflected in the weak growth of the inverse
	Knudsen number shown in Fig.~\ref{F1}a.}
Comparing with the conformal case $\Delta{\,=\,}1$ where $\bar\pi$ drops to ${\approx\,}0$ after several dozen fm/$c$ \cite{Jaiswal:2019cju} one realizes that increasing the growth rate for the mean free path by raising $\Delta$ from 1 to 3 strongly delays thermalization.

As already noted, the inverse Knudsen number shown in Fig.~\ref{F1}a increases with time as a power law $\bar\tau{\,\sim\,}\tau^\alpha$ with $\alpha{\,\simeq\,}1/4$. This can be understood as follows: after rapid decay of the initial deviation of $\bar\pi$ from the attractor value (after a time $\tau{\,\lesssim\,}0.1 \tau_\pi$, i.e.\ at Kn${\,\gtrsim\,}10$, {\it cf.}\ Fig.~\ref{F1}c), the shear stress follows the attractor which for $\Delta{\,=\,}3$ remains for a long time close to the free-streaming value of 1/4. For $10{\,\lesssim\,}\tau/\tau_0{\,\lesssim\,}10^3$ the evolution of the inverse Knudsen number can therefore be solved approximately by using Eq.~(\ref{Knudsen}) with $\Delta{\,=\,}3$ to yield $\bar\tau{\,\sim\,}\tau^{\bar\pi_\mathrm{attr}}$. The slightly larger slope in Fig.~\ref{F1}a for aHydro (dashed blue) compared to CE (solid red) reflects the slightly larger attractor value for $\bar\pi$ in aHydro.

%
%

%
%

The increase with time of Kn$^{-1}$ is halted in aHydro by further increasing $\Delta$ to 4 --- see  Fig.~\ref{F2}a. This is again understood by writing Eq.~(\ref{Knudsen}), after decay of the initial deviation from the attractor at some time $\tau_*$, as
\begin{align}
\label{pibarevol}
   \bar\tau \approx \bar\tau_* \left(\frac{\tau}{\tau_0} \right)^{1 + \Delta (\bar\pi_\mathrm{attr} - 1)/3},
   \qquad \bar\tau_* \equiv \bar\tau(\tau_*).
\end{align}
Taking $\bar\pi^\mathrm{attr}{\,=\,}\frac{1}{4}$ for aHydro, the effects of viscous heating should be exactly balanced by the now even faster growth of the mean free path, leading to constant inverse Knudsen number. However, owing to the weak coupling between the evolutions of $\bar\pi$ and $\bar\tau$, $\bar\pi$ decreases very slightly with time, leading to $\bar\pi^\mathrm{attr}(\tau){\,<\,}\bar\pi^\mathrm{attr}_{0,\mathrm{aHydro}}{\,=\,}\frac{1}{4}$ at any time $\tau{\,\ne\,}0$. We checked that this explains quantitatively the slightly negative slope of the aHydro attractor in Fig.~\ref{F2}a, as well as the larger slope for CE since $\bar\pi^\mathrm{attr}_{0,\mathrm{CE}}{\,<\,}\bar\pi^\mathrm{attr}_{0,\mathrm{aHydro}}$. On much longer time scales Kn$^{-1}$ is numerically found to continue its decrease, never thermalizing but approaching asymptotic free-streaming. Accordingly, the inverse Reynolds number shown in Fig.~\ref{F2}b never drops visibly below 1/4 and approaches exactly that value at late times.
For $\Delta=5$ (Fig.~\ref{F3}) the inverse Knudsen number Kn$^{-1}$ decreases more rapidly with time, and the system moves away from thermalization more quickly. When initialized in thermal equilibrium, $\bar\pi_0=0$, the inverse Reynolds number quickly approaches the free-streaming value and stays there (Fig.~\ref{F3}b). Plotting the inverse Reynolds number against the inverse Knudsen number in Fig.~\ref{F3}c, early times now correspond to ``large'' Kn$^{-1}$ near the right edge of the plot, and the system evolves leftward towards smaller values of Kn$^{-1}$, all at constant Re$^{-1}=\bar\pi_0^\mathrm{attr}$. 

%
\begin{figure}[b!]
\vspace*{-5mm}
 \includegraphics[width=\linewidth]{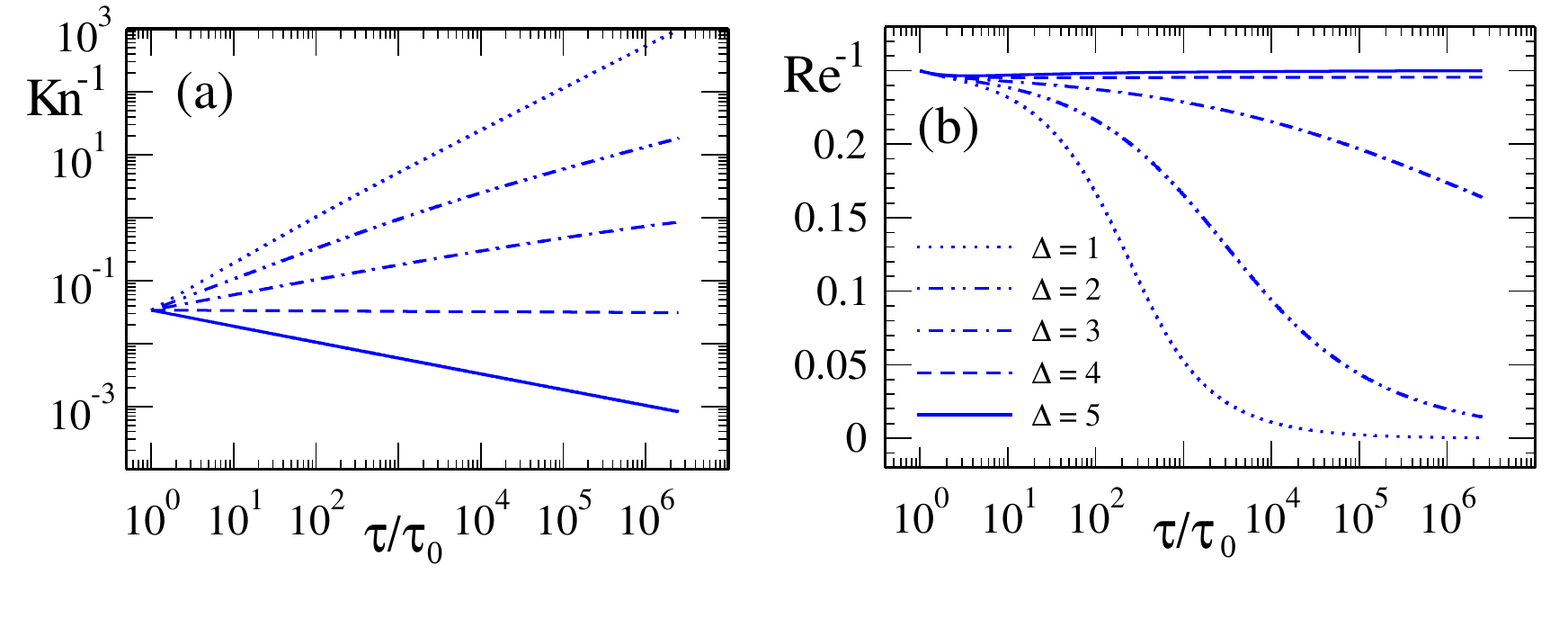}
 \vspace{-6mm}
 \caption{(Color online) Time evolution of Kn$^{-1}{\,=\,}\bar\tau$ (panel a) and 
	Re$^{-1}{\,=\,}\bar\pi$ (panel b) for different values of $\Delta$ obtained using aHydro 
  	with initial conditions on the hydrodynamic attractor.
	\label{F4}}
\end{figure} 
%

We conclude this part of our discussion by summarizing in Fig.~\ref{F4} the time evolution of the attractor solutions for the inverse Knudsen and Reynolds numbers for $\Delta=1$ through 5. For $\Delta{\,<\,}4$ the attractors are characterised by Kn$^{-1}$ increasing with time, although the growth keeps falling as $\Delta$ gets bigger, delaying thermalization. For $\Delta{\,=\,}4$, Kn$^{-1}$ begins to decrease with time; as $\Delta$ increases further this trend strengthens. The shear stress shown in Fig.~\ref{F4}b decreases with time for $\Delta{\,=\,}1$, indicating thermalization; thermalization is delayed for $\Delta{\,=\,}2,\, 3$, and it never happens for $\Delta{\,\geq\,}4$ where $\bar\pi$ is approximately frozen at the free-streaming value.

%
\begin{figure}[b!]
\begin{center}
\hspace*{-4mm}
    \includegraphics[width=1.05\linewidth]{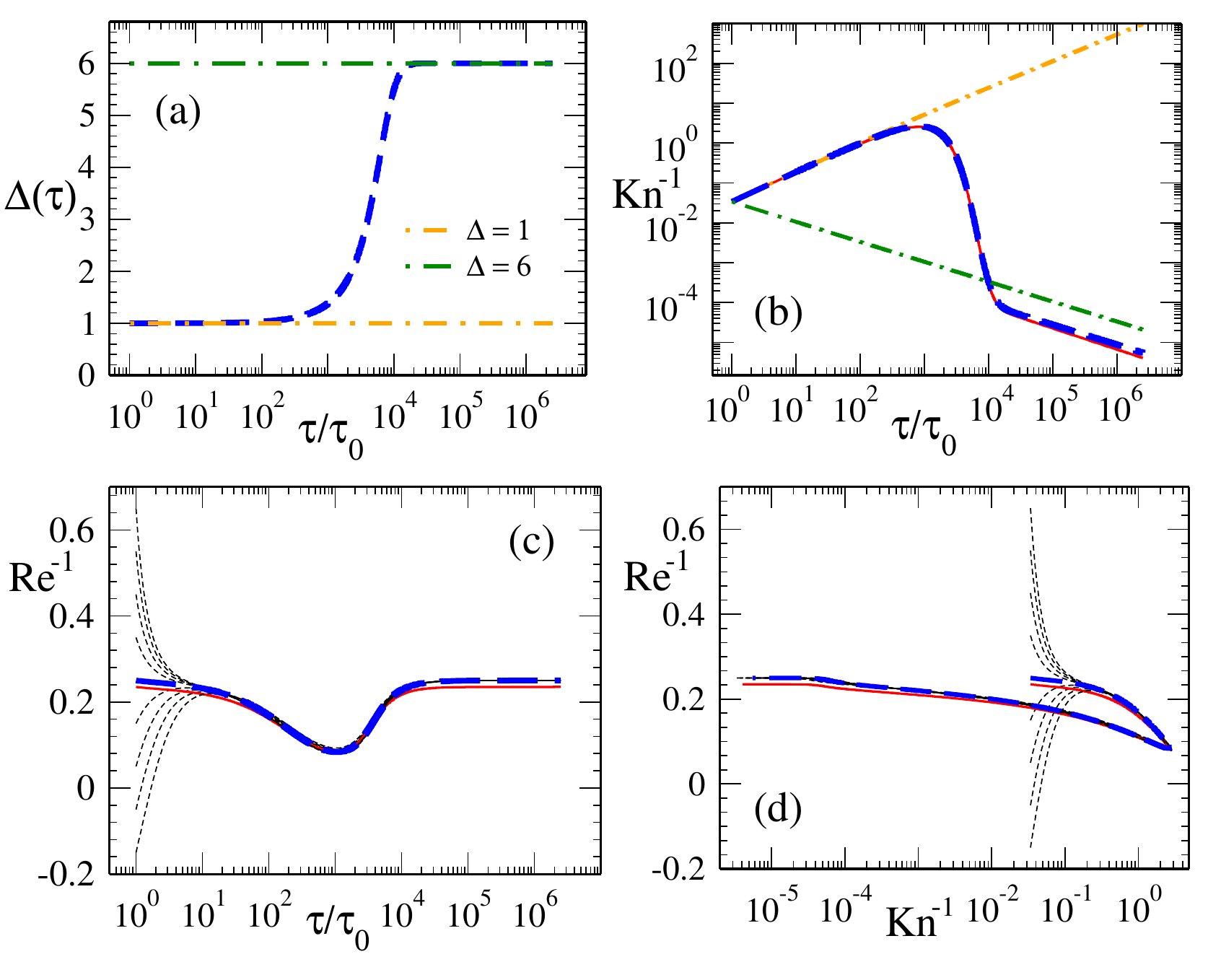}
\end{center}
\vspace{-4mm}
\caption{(Color online) 
  	(a): Parametrisation of $\Delta(\tau)$. (b) and (c): time evolution of Kn$^{-1}{\,=\,}\bar\tau$ 
	and Re$^{-1}{\,=\,}\bar\pi$, respectively. (d): evolution of Re$^{-1}$ as a function of Kn$^{-1}$.
	Note the change of direction of the arrow of time which starts out pointing right but then turns 
	around to point left at late times $\tau{\,>\,}\tau_\mathrm{tr}$. As before, the thick dashed blue
	and thin solid red lines are aHydro and CE attractors, respectively. Black thin dashed lines are 
	from aHydro with off-attractor initial conditions. 
         \label{F5}}
\end{figure}  
%

\noindent {\it (b)\,Time-dependent $\Delta(\tau)$.}
To simulate within the restrictions of Bjorken symmetry the transition from early thermalizing to late decoupling dynamics, and to explore the behavior of the hydrodynamic attractor during this transition, we study a time-dependent $\Delta(\tau)$ that interpolates between $\Delta_i{\,=\,}1$ at early and $\Delta_f{\,=\,}6$ at late times, with a transition around $\tau_\mathrm{tr}{\,=\,}10\,\mathrm{fm}{\,=\,}2500\, \tau_0$ ({\it cf.}~Fig.~\ref{F5}a):\footnote{%
	Note that now the last term in Eq.~(\ref{Knudsen}) couples the evolution of the Knudsen 
	number directly to that of the energy density (for constant $\Delta$ its evolution was decoupled).
	}
\begin{equation}
\label{Deltatau}
   \Delta(\tau) = \frac{\Delta_{f} \Delta_{i}}
                                {\Delta_i + (\Delta_{f}{-}\Delta_{i})\,e^{-(\tau{-}\tau_0)/\tau_\mathrm{tr}}}
                                \qquad \text{for} \ \ \tau\geq\tau_0.
\end{equation}
Figure~\ref{F5} shows that even with this time-dependent $\Delta$ the evolution of the inverse Reynolds number is controlled by a hydrodynamic attractor. At early times $\tau{\,<\,}\tau_\mathrm{tr}$ the inverse Knudsen number Kn$^{-1}$ increases with time, i.e. the system moves closer to local equilibrium. Around $\tau{\,\sim\,}\tau_\mathrm{tr}$, Kn$^{-1}$ drops steeply by almost 3 orders of magnitude and then continues to decrease more moderately, following the attractor for fixed $\Delta_f{\,=\,}6$ and thereby moving rapidly away from thermal equilibrium. Fig.~\ref{F5}c shows that the inverse Reynolds number quickly relaxes from its initial condition to the attractor for $\Delta_i{\,=\,}1$, starting close to the free-streaming limit but decreasing slowly for $\tau{\,<\,}\tau_\mathrm{tr}$ as the system moves towards local equilibrium. This decrease is halted at $\tau{\,\sim\,}\tau_\mathrm{tr}$ after which Re$^{-1}$ increases again and soon settles on the free-streaming limit (Re$^{-1}_\mathrm{fs}=0.25$ for aHydro and ${\approx\,}0.235$ for third-order CE). Plotting the evolution of Re$^{-1}$ against the inverse Knudsen number in Fig.~\ref{F5}d we see the system initially moving right and slightly downward before turning around and moving left again and slightly upward. This is qualitatively reminiscent of the pattern shown in Fig.~2 of Ref.~\cite{Romatschke:2017acs} for (2+1)-dimensional hydrodynamic expansion with longitudinal boost-invariance. We checked with additional calculations that the system joins the free-streaming attractor similarly rapidly if the evolution starts away from the attractor at a time $\tau_i{\,\gg\,}\tau_0{\,=\,}0.004$\,fm/$c$ (for example at $\tau_i{\,=\,}\tau_\mathrm{tr}$ or $\tau_i{\,=\,}100\,\tau_\mathrm{tr}$) as long as the initial Knudsen number is large, $\bar\tau_0 \ll 1$.  

\begin{figure}[h!]
\begin{center}
\hspace*{-4mm}
    \includegraphics[width=0.8\linewidth]{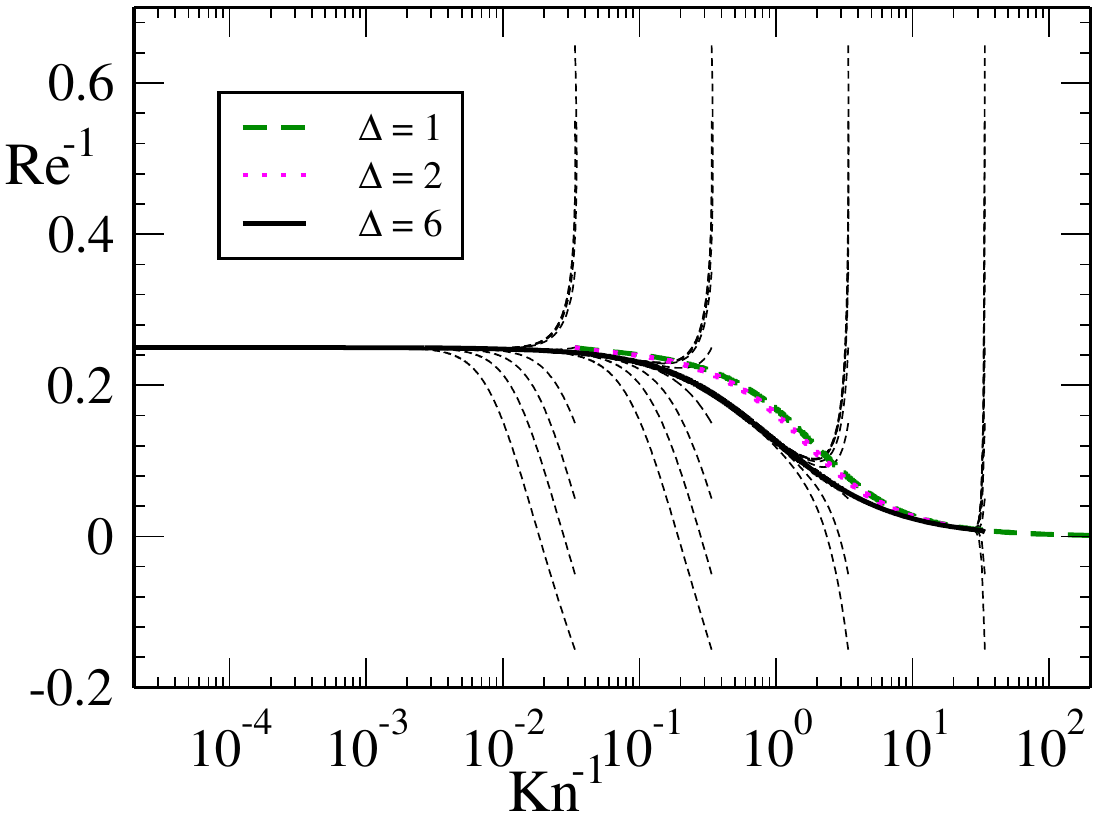}
\end{center}
\vspace{-4mm}
\caption{(Color online) 
	The aHydro attractor for Re$^{-1}$ for $\Delta{\,=\,}1$ (dashed green), $\Delta{\,=\,}2$ 
	(dotted magenta), and $\Delta{\,=\,}6$ (solid black), together with a few solutions for 
	$\Delta{\,=\,}6$ with off-attractor initial conditions (thin black dashed) imposed at various 
	initial values for the inverse Knudsen number Kn$^{-1}$.   	
         \label{F6}}
\end{figure}

Figure~\ref{F6} shows (as dashed green, dotted magenta and solid black lines for $\Delta{\,=\,}1$, 2, and 6, respectively) the aHydro attractor for Re$^{-1}$ as a function of the inverse Knudsen number Kn$^{-1}$, together with a number of dashed lines indicating aHydro solutions with various off-attractor %
%
initial conditions. One should note the characteristic change of shape of the thin dashed lines as the Knudsen number increases from the right towards the left, reflecting the transition near Kn${\,\approx\,}1$ from exponential to power-law decay of initial deviations of Re$^{-1}$ from the attractor value \cite{Jaiswal:2019cju, Kurkela:2019set}. Even though the change of the temperature dependence of the relaxation time $\tau_\pi$ between $\Delta{\,=\,}1$ and $\Delta{\,=\,}6$ reflects a significant change in a key medium property of the fluid, the attractors for these differing $\Delta$-values remain rather close to each other. While for $\Delta{\,=\,}1$ and 2 the system evolves in time from the left towards the right along the corresponding attractor, the arrow of time points in the opposite direction (i.e.\ from the right towards the left) for $\Delta{\,=\,}6$. Comparison with Fig.~\ref{F5} clarifies how in Fig.~\ref{F5}d the system first moves right along the dashed green attractor for $\Delta{\,=\,}1$ and then crosses over to the solid black $\Delta{\,=\,}6$ attractor, turning around in the process to follow the latter towards the left.
  
\textit{\textbf{4.\,Summary.}} We found that by parametrically modifying the temperature dependence of the relaxation time in a simple (0+1)-dimensional expanding fluid one can systematically study the transition from thermalizing dynamics for small and decreasing Knudsen numbers to decoupling and eventual free-streaming dynamics at large and increasing Knudsen numbers. In realistic (3+1)-dimensional simulations of heavy-ion collisions this transition is caused by the change in dimensionality of the expansion, from initially (0+1)-d to eventually (3+1)-d, and can only be studied numerically. In our (0+1)-d Bjorken toy model we trigger this transition by modifying the temperature dependence of the microscopic scattering rate as a function of time. This allows us to study this transition in significant detail and with a large degree of semi-analytic control. We find that throughout its history, from initial thermalizing dynamics through the transition towards dynamical decoupling and ultimately free-streaming dynamics, the system's evolution is controlled by a hydrodynamic attractor. We computed this attractor for relativistic anisotropic and 3$^\mathrm{rd}$-order Chapman-Enskog dissipative hydrodynamics and discussed its properties. The shape of the attractor for Re$^{-1}$ as a function of Kn$^{-1}$ was found to exhibit only weak sensitivity to changes in the medium properties distinguishing thermalizing from decoupling dynamics. The only major distinction between these different dynamics is the direction in which the expanding system evolves along this quasi-universal attractor. Up to necessary redefinitions of the Knudsen and inverse Reynolds numbers to account for transverse expansion and the existence of multiple shear stress components together with a bulk viscous pressure, we expect a similar attractor to rule all fluid cells in a three-dimensionally expanding fluid, with the possible exception of dilute cells near its outer surface whose microscopic relaxation times may be too long for them to reach the attractor before the fluid breaks up into free-streaming hadrons. Future work will check this expectation.

\textit{\textbf{Acknowledgements:}}
The authors thank G. Denicol, M. Heller, S. Jaiswal, J. Noronha, P. Romatschke and V. Svensson for fruitful discussions and insightful comments. This work was supported by the U.S. Department of Energy, Office of Science, Office for Nuclear Physics under Award No. \rm{DE-SC0004286}.

\bibliographystyle{elsarticle-num}
\bibliography{references}

\end{document}